\documentclass[twocolumn,preprintnumbers,amsmath,amssymb]{revtex4}

\usepackage{graphicx}
\usepackage{dcolumn}
\usepackage{bm}

\begin{document}
\title{On the Possibility of an Electronic-structure Modulation Transistor}
\author{Hassan Raza}
\affiliation{School of Electrical and Computer Engineering, Cornell University, Ithaca, NY, 14853}
\author{Tehseen Z.~Raza}
\affiliation{School of Electrical and Computer Engineering, West Lafayette, IN, 47907}
\author{Tuo-Hung Hou}
\affiliation{Department of Electronics Engineering, National Chiao Tung University, Hsinchu, Taiwan 300, ROC.}
\author{Edwin C. Kan}
\affiliation{School of Electrical and Computer Engineering, Cornell University, Ithaca, NY, 14853}

\begin{abstract}
We present a novel electronic-structure modulation transistor (EMT), which can possibly be used for post-CMOS logic applications. The device principle is based on the bandwidth modulation of a midgap or near-midgap localized state in the channel by a gate voltage. A single-band tight-binding method coupled with non-equilibrium Green's function formalism for quantum transport is employed to predict the IV characteristics. Our objective is to confirm if an EMT has a self gain and if it can overcome the 2.3kT/decade thermal limit with low supply voltage. The ON current depends on the bandwidth of the state and is limited by the quantum of conductance for a single band. The OFF current is set by the gate leakage and tunneling through the higher bands, which is expected to be small if these bands are a few eV above the energy level of the localized state. 
\end{abstract}

\maketitle

\section{Introduction and EMT concept}
Electronic-structure modulation has not yet been realized in practical devices. We envision that transistors based on the electronic-structure modulation of the channel material can open up a new class of post-CMOS logic devices. Electronic-structure modulation transistor (EMT) can be achieved in one, two or three dimensional channels. In contrast to the conventional field-effect transistor (FET) that rely on the band edge shift using an external gate voltage, in EMT, the gate voltage modifies the band dispersion and hence modulates the bandwidth. FETs are thus limited by the 2.3kT/decade limit in their subthreshold inverse slope \cite{Taur}, where k is the Boltzmann constant and T is the temperature. With the scaling of the supply voltage, the OFF channel current is increasing \cite{Taur}, making the power dissipation a show stopper. It is therefore desirable to reduce the OFF current, even with a low supply voltage, while retaining the self gain. 

Concepts based on the modulation of various device parameters have been explored earlier. For example, velocity/mobility modulation transistors rely on the real-space transfer of carriers between two adjacent materials with different mobilities \cite{Sze}. Similarly, quantum modulation transistors (QMT) are based on the constructive and destructive interference of the wavefunctions in the channel by electrically changing the T-shaped box dimensions \cite{Hess89}. Furthermore, quantum effects in various planar heterostructures based on the modulation-doped field-effect transistor (MODFET) principle have been explored \cite{Ismail91}, where the field-effect is used to perturb the barriers for carriers flowing between the source and the drain electrodes. The localization of the state near band edges due to disorder in the Anderson localization is also a relevant concept, which leads to a mobility edge \cite{Barnham01}, but this effect is limited by the thermal limit.

\begin{figure}[!t]
\centering
\vspace{2.0in}
\hskip-2.5in\includegraphics{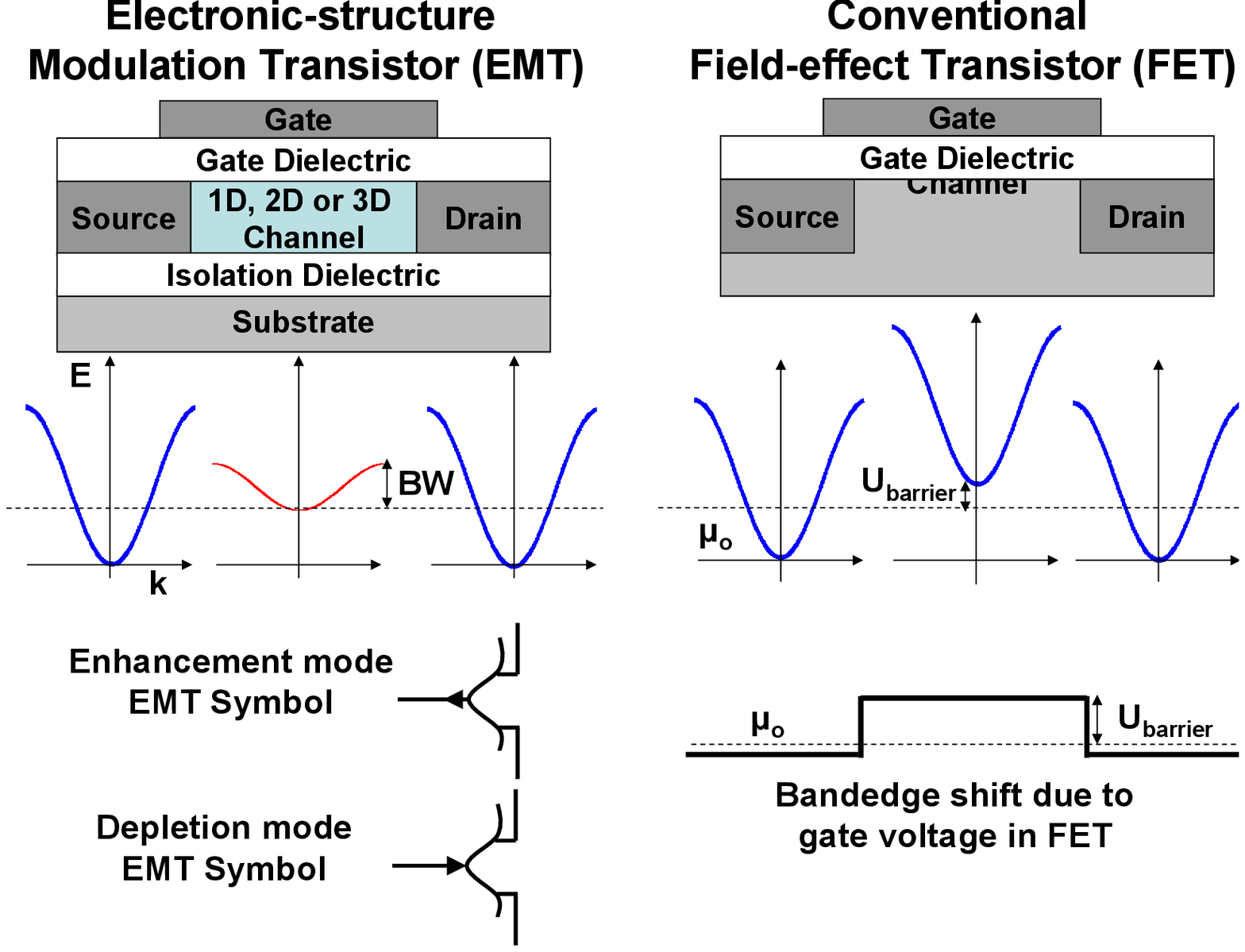}
\caption{Device principle of an EMT. The conventional field-effect transistor depend on the gate voltage dependent shift of the bandedge giving rise to the barrier potential ($U_{barrier}$), resulting in thermal limit of 2.3kT/decade in inverse subthreshold slope. For EMT, the bandwidth (BW) is dependent on the gate voltage and results in modulation of channel transport. EMT device structure is shown as well. Proposed symbols for enhancement mode and depletion mode EMT are shown.}
\label{fig_3}
\end{figure}

In this work, we propose a novel EMT with a few kT of supply voltage, whose working principle is based on the modulation of the bandwidth of a state by using external gate voltage. It has been suggested that above a certain tip voltage in the scanning tunneling spectroscopy, the coupling between otherwise localized highest occupied molecular orbital energy levels of the styrene molecules in a one dimensional chain on H:Si(100)-(2$\times$1) surface increases \cite{Raza08_prb_1,Wolkow05}. It is equivalent to changing the hopping integral between the two styrene molecules acting as lattice sites, which in turn changes the bandwidth of the localized state. Such electronic-structure modulation of bandwidth can not only be achieved by changing the coupling between lattice sites, but also by taking advantage of the k-dependent and gate voltage dependent real-space localization of a state. We present one such example of a N=9 zigzag graphene nanoribbon to demonstrate the concept using a simple $p_z$-orbital model with a fixed tight-binding parameter. Furthermore, electronic-structure modulation has also been reported in graphene nanoribbons \cite{Raza08_ac_prb,Novikov07} and bilayer graphene \cite{Raza_Bilayer,McCann06}. In this work, we have developed the necessary theory to analyze the switching operation of EMT.

\begin{figure}[!t]
\centering
\vspace{1.8in}
\hskip-3.0in\includegraphics{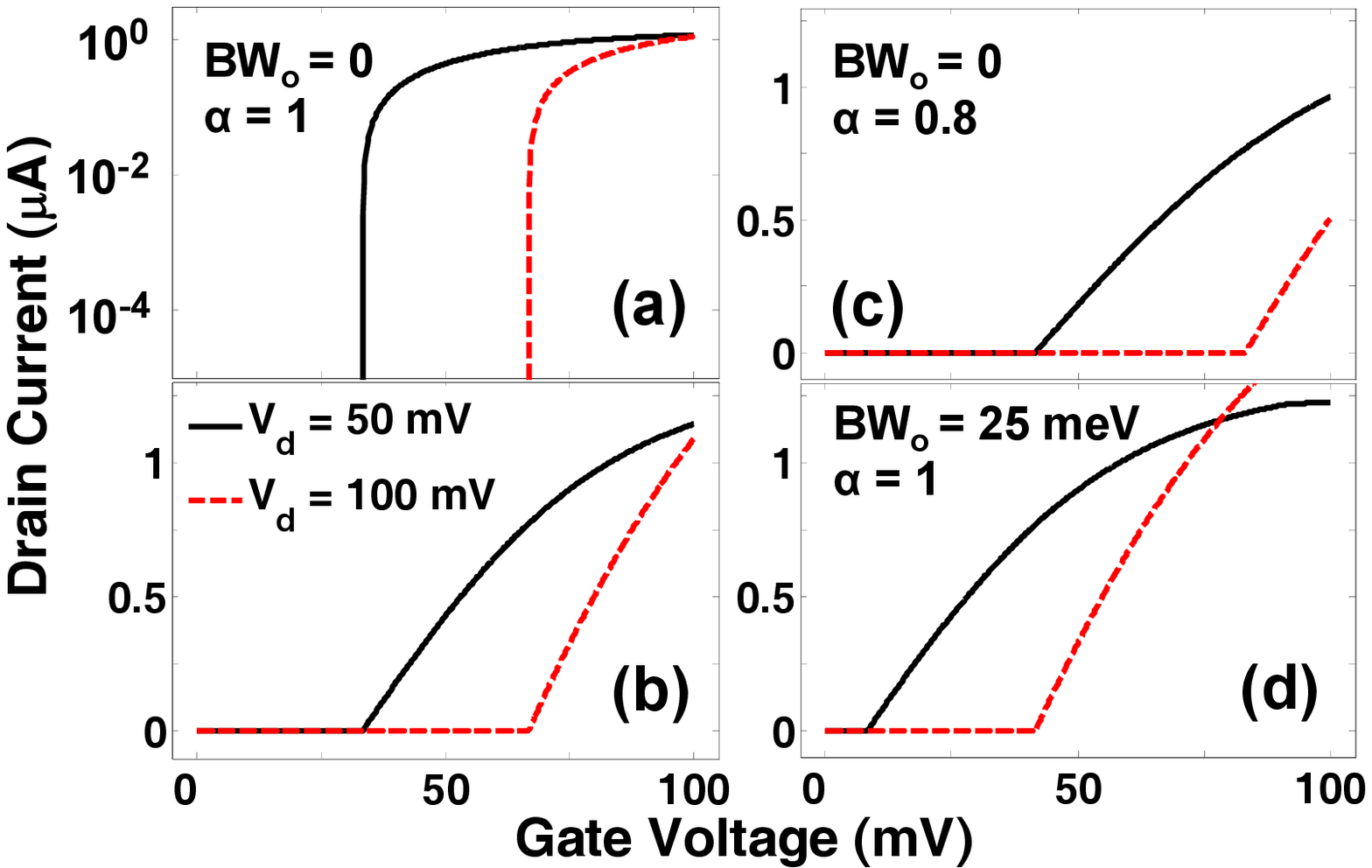}
\caption{Transfer characteristics of an EMT. (a,b) Transfer characteristics for $\alpha$ = 1 and $BW_o$ = 0 on a logarithmic and a linear scale, showing large ON/OFF current ratio. (c) Transfer characteristics for $\alpha$ = 0.8 and $BW_o$ = 0, showing that the transfer curves are shifting towards the higher gate voltage. (d) Transfer characteristics for $\alpha$ = 1 and $BW_o$ = 25 meV, depicting a shift towards the lower gate voltage.}
\label{fig_1}
\end{figure}

\section{Theoretical Model}

We start with the following ansatz for the bandwidth:
\begin{eqnarray}BW = Mag(\alpha |eV_g | + BW_o) \end{eqnarray} 
where $\alpha$ is a dimensionless parameter, which we call the modulation factor. $BW_o$ is the residual BW at zero gate voltage. A positive $\alpha$ results in enhancement mode characteristics, whereas a negative $\alpha$ may give either depletion or enhancement mode characteristics depending on the value of $BW_o$. The BW is zero for a channel with midgap or near midgap localized states. No channel current flows in such a case, except for the dielectric leakage current and tunneling through the higher bands, which should be couple of eV above the localized state to minimize the leakage. By applying a gate voltage to increase the bandwidth of the state, the current starts to flow. Furthermore, the modulation behavior may very well be nonlinear in various materials, which can be readily incorporated in our model. 

Within the single-band tight-binding (SBTB) approximation \cite{Kittel_Book} with the nearest-neighbor coupling or hopping parameter of $t_o$, the dispersion relationship for a one dimensional channel is given as:
\begin{eqnarray}E(k)=E_b+2t_o[1-cos(ka)]\end{eqnarray}  
In such a dispersion, the BW is $| 4t_o |$, where $E_b$ and $E_b+4t_o$ are the lower and upper band edges, respectively, for a positive $t_o$ and \textit{vice versa} for a negative $t_o$. Using Eq. 1, $t_o$ is thus gate voltage dependent as:
\begin{eqnarray}t_o = \alpha |eV_g |/4 + BW_o/4 \end{eqnarray}
The SBTB Hamiltonian with five lattice points is given as:
\begin{eqnarray}H=\begin{cases}E_{b}+2t_o+U_L(i,j)\ \ \ for\ i=j\cr -t_o\ \ \ for\ |i-j|=1;\ i,j = 1,2,..,5 \end{cases}\end{eqnarray}
The Laplace potential ($U_L$) due to the drain bias ($V_d$) is included as a linear voltage drop. The lower band edge ($E_b$) is taken to be equal to the equilibrium chemical potential. Hartree potential is ignored for simplicity, since it does not affect the device operation, although it may affect the numerical results. 

\begin{figure}[!t]
\centering
\vspace{1.8in}
\hskip-3.0in\includegraphics{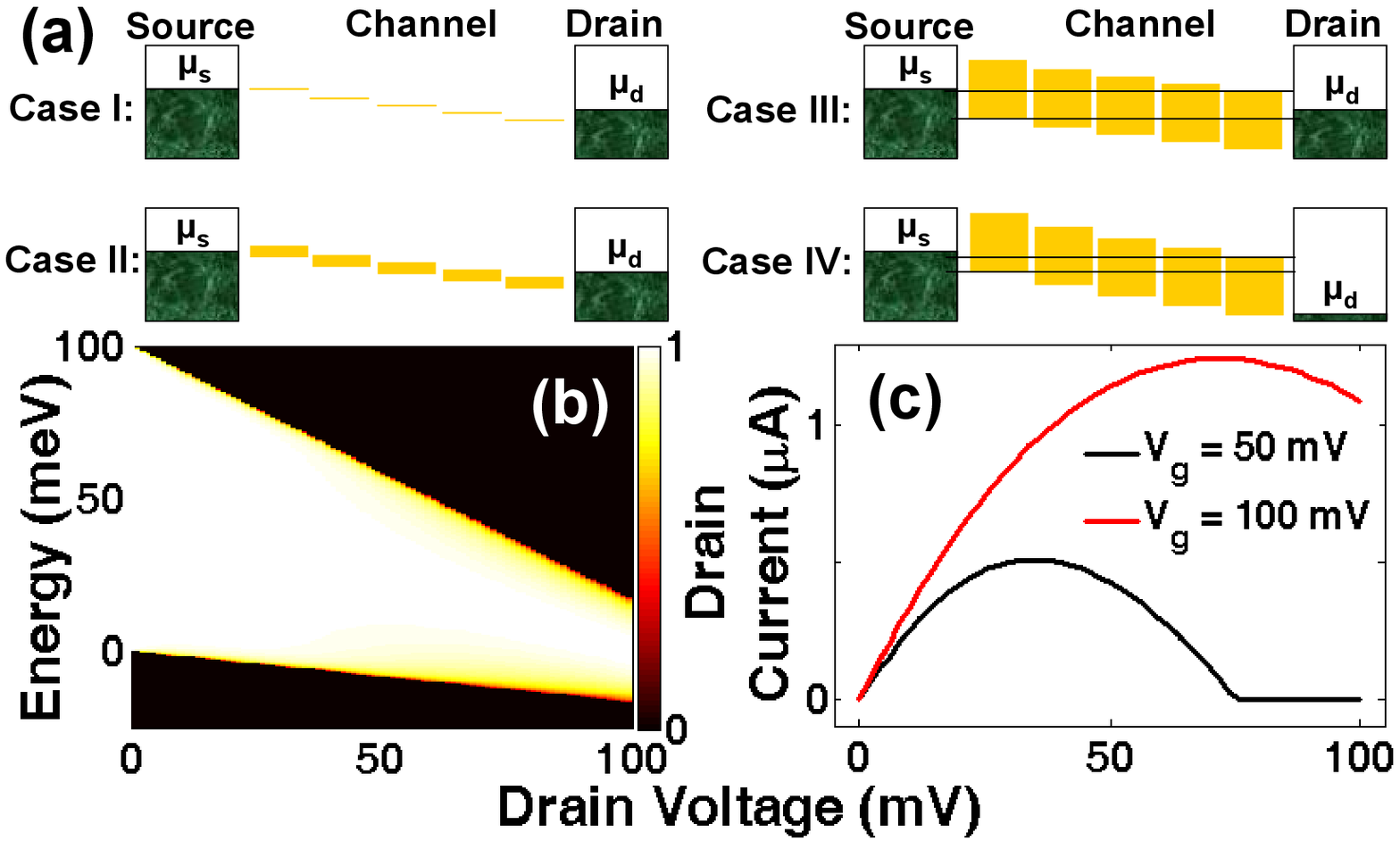}
\caption{Output characteristics of an EMT. (a) Channel conduction window - Case I: small bandwidth does not result in a conduction window between source and drain; Case II: even a slightly larger bandwidth is not sufficient; Case III: a conduction window is achieved for a larger bandwidth; Case IV: a further increase in drain bias results in a small conduction window for the same bandwidth as in case III. (b) Decreasing transmission window with increasing drain bias for $V_g$ = 100 mV, $\alpha$ = 1 and $BW_o$ = 0. Transmission window at zero drain bias is 100 meV, as expected. (c) Output characteristics showing negative differential conductance for $\alpha$ = 1 and $BW_o$ = 0. The inflexion point is at a higher drain bias for a higher gate voltage.}
\label{fig_2}
\end{figure}

We use the non-equilibrium Green's function formalism (NEGF) for calculating the quantum transport \cite{Datta05}, though we consider only the coherent limit where it is equivalent to the Landa\"uer's approach and the current can be evaluated from the transmission as:
\begin{eqnarray}I_d=2\ (for\ spin)\times\frac{q}{h}\int{dE\ T(E) [f_s-f_d]}\end{eqnarray}
where transmission is $T(E)=tr(\Gamma_{1}G\Gamma_{2}G^{\dagger})$. The Green's function for the channel is,
\begin{eqnarray}G=[(E+i0^+)I-H-\Sigma_{1,2}]^{-1}\end{eqnarray}
Self-energies and broadening functions are $\Sigma_{1,2}=-t_oe^{ik_{1,2}a}$ and $\Gamma_{1,2}=i(\Sigma_{1,2}-\Sigma_{1,2}^\dagger)$, respectively. $f_{s,d}=[1+e^{(E-\mu_{s,d})/kT}]^{-1}$ are the contact Fermi functions. $\mu_{s,d}$ are source/drain Fermi energies. $\mu_d$ is shifted due to drain bias as $\mu_d = \mu_o - qV_d$ and  $\mu_s = \mu_o$, where $\mu_o$ is the equilibrium chemical potential. 

\section{Discussion of Results}

We first discuss the theoretical results for a device with $\alpha$ = 1 and $BW_o$ = 0. The transfer characteristics of an EMT with $V_d$ = 50 mV and 100 mV are shown in Figs. 2(a,b) on a logarithmic and a linear scale, respectively. The transconductance is about 0.05 mS. As shown, a large difference in the ON/OFF current is obtained even for a small gate voltage. We further show the transfer characteristics for a smaller $\alpha$ = 0.8 in Fig. 2(c), which results in a delayed onset of conduction. For $BW_o$ = 25 meV and $\alpha$ = 1, we illustrate the transfer characteristics in Fig. 2(d). A finite $BW_o$ results in shifting the onset of conduction towards the lower gate voltage due to a higher total BW across channel. 

The Laplace potential due to the drain bias can block the conduction under coherent conditions and delay the onset of conduction as shown in Fig. 2 and schematically explained in Fig. 3(a). Apart from  the dielectric leakage current, zero BW results in a zero channel current. However, an incremental gate voltage, only slightly increasing the BW, would not lead to any transport because the drain bias also shifts the energy windows linearly for various lattice points. A conduction window thus cannot be established as shown in Fig. 3(a) for case I and case II. Substantial gate voltage is required to make a conduction window between the source and the drain as shown in case III in Fig. 3(a). 

In coherent transport, the lower limit of transmission as a function of the drain bias is due to the lower band edge shift of the lattice point closer to the source. Similarly the upper limit is due to the upper band edge shift of the lattice point closer to the drain as shown in Fig. 3(a). A higher drain bias reduces this conduction window as schematically shown in case IV of Fig. 3(a). The decreasing transmission window and the increasing Fermi function difference lead to increasing and then decreasing $I_d$ as shown in Fig. 3(c). The output resistance strongly depends on the load line, the highest limit is infinity at the inflexion point. Therefore, an EMT can have a self gain much larger than unity. To obtain ps operation for the reported characteristics, the total EMT capacitance for calculating the \emph{CV/I} should be less than 10 aF, which is practically realizable.

We now consider the case of a zigzag graphene nanoribbon (zzGNR) using a $p_z$-orbital tight-binding model with a parameter of 2.5 eV, schematically shown in Fig. 4(a). It has been reported that a N = 9 zzGNR has a dispersion-less midgap trap state \cite{Liang07} as shown in Fig. 4(b). We report that we can modulate the bandwidth of this midgap state by applying an electric-field (E) in the width direction with gate voltage applied at the open-ring end of the zzGNR as shown in Fig. 4(b) with E = 1 V/nm. At the X point, the midgap state is perfectly localized at the carbon atoms close to the Ground in Fig. 4(c). Therefore, this state does not shift with an electric-field at X point. Whereas, at $\Gamma$ point, the state is distributed on the odd-numbered atoms closer to the open benzene ring in Fig. 4(c). A positive gate voltage results in an upward shift of energy values at the $\Gamma$ point, resulting in the dispersion of the midgap state as shown in Fig. 4(b). We report the bandwidth dependence on the gate voltage in Fig. 4(d), which follows Eq. 1 with $\alpha$ $\approx$ 0.9 and $BW_o$ = 0, provided the gate voltage drops linearly across the GNR only - such an ideal case possibly achievable by using high-K insulators. The higher bands are also expected to lead to additional source-drain leakage current apart from the gate leakage current, which is expected to be small because these bands are almost 2 eV above the midgap state. Although, a simple $p_z$-orbital model may or may not give the correct ground state properties for this zzGNR, nonetheless it serves as a good example to the demonstrate the electronic-structure modulation principle of a dispersionless state. 

\begin{figure}[!t]
\centering
\vspace{1.8in}
\hskip-2.5in\includegraphics{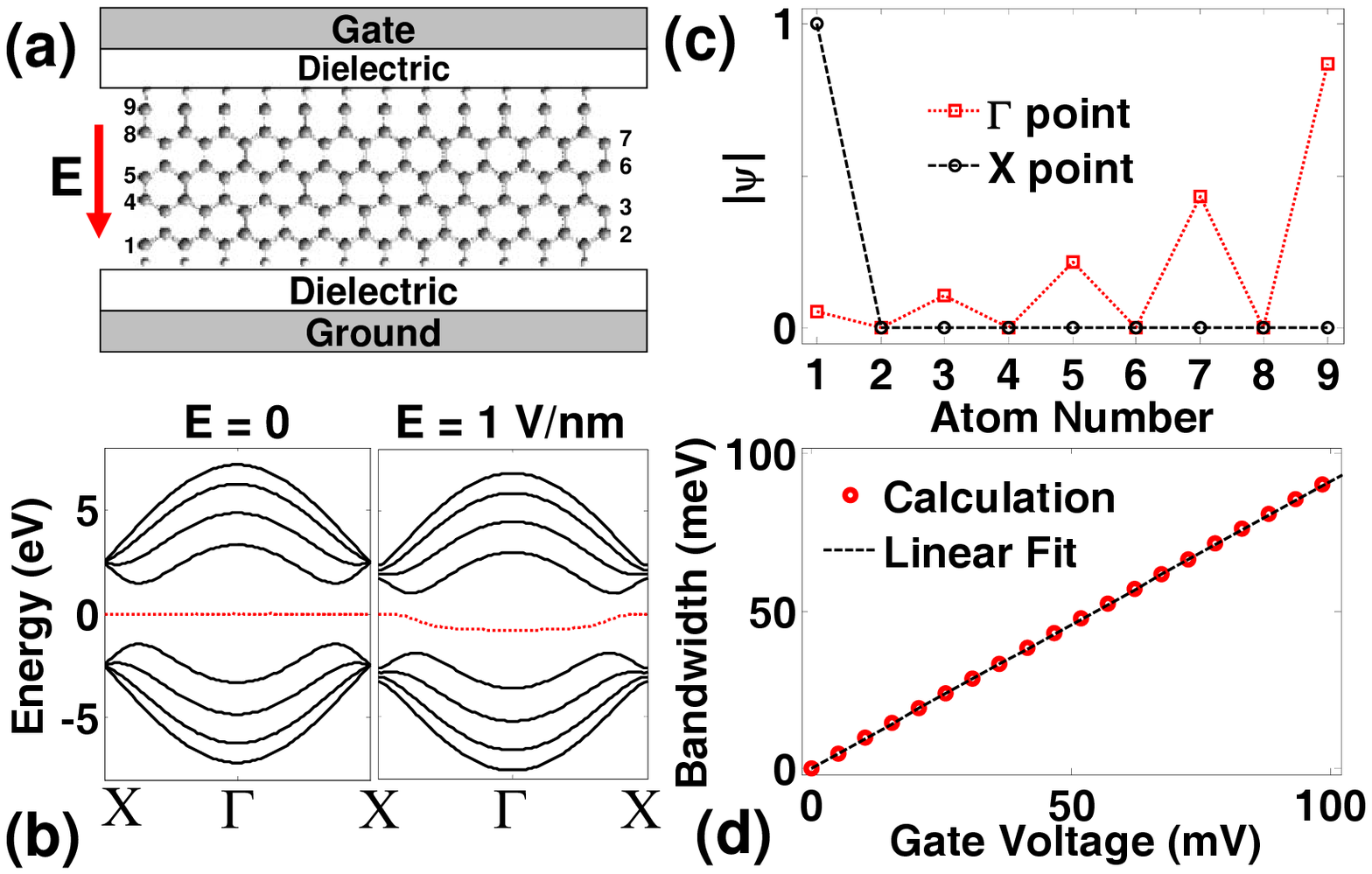}
\caption{Electronic structure modulation of zzGNR. (a) The schematics of N = 9 zzGNR. (b) For X-point, the wavefunction is localized on the first layer of atoms towards ground contact. For $\Gamma$-point, the wavefunction is majorly localized on odd-numbered layers towards open benzene rings edge. (c) A midgap trap state is observed without an electric-field (E). An electric-field results in some dispersion of the midgap state. (d) Bandwidth is linearly dependent on the gate voltage, yielding $\alpha$ $\approx$ 0.9 and $BW_o$ = 0. Atomic visualization was done using GaussView \cite{GW03}.}
\label{fig_5}
\end{figure}

\section{Conclusions}

We have theoretically analyzed an EMT with few kT supply voltage based on modulating the electronic-structure of a channel by the external gate voltage. We have developed a theoretical framework based on SBTB method coupled with NEGF to analyze the characteristics of EMT. We report that one can obtain self gain and large ON/OFF channel current ratio with few kT supply voltage. 

\section*{Acknowledgment}

H. Raza is thankful to S. Datta, M. A. Alam and D. Stewart for useful discussion. The work is supported by NSF and by NRI through CNS at Cornell University, Ithaca, NY.

\end{document}